\documentclass{vgtc}                
\ifpdf
  \pdfoutput=1\relax                   
  \usepackage{graphicx}                
  \DeclareGraphicsExtensions{.pdf,.png,.jpg,.jpeg} 
\else
  \usepackage{graphicx}                
  \DeclareGraphicsExtensions{.eps}     
\fi%

\graphicspath{{figures/}{pictures/}{images/}{./}} 

\usepackage{microtype}                 
\PassOptionsToPackage{warn}{textcomp}  
\usepackage{textcomp}                  
\usepackage{mathptmx}                  
\usepackage{times}                     
\usepackage{cite}                      
\usepackage{tabu}                      
\usepackage{booktabs}                  
\usepackage{wrapfig}
\usepackage{xcolor}

\usepackage{hyperref}

\onlineid{1038}

\vgtccategory{Research}
\vgtcinsertpkg

\title{Interactive Probing of Multivariate Time Series Prediction Models: A Case of Freight Rate Analysis}

\author{Xu Haonan\thanks{e-mail: haonan.xu.2020@msc.smu.edu.sg}\\ %
        \scriptsize Singapore Management University %
\and Li Haotian\thanks{e-mail: haotian.li@connect.ust.hk}\\ %
     \parbox{2.5in}{\scriptsize \centering Hong Kong University of Science and Technology \\ Singapore Management University}%
\and Wang Yong\thanks{e-mail: yongwang@smu.edu.sg}\\ %
     {\scriptsize Singapore Management University}}

\abstract{
We present an interactive probing tool to create, modify and analyze what-if scenarios for multivariate time series models. The solution is applied to freight trading, where analysts can carry out sensitivity analysis on freight rates by changing demand and supply-related econometric variables and observing their resultant effects on freight indexes. We utilize various visualization techniques to enable intuitive scenario creation, alteration, and comprehension of time series inputs and model predictions. Our tool proved to be useful to the industry practitioners, demonstrated by a case study where freight traders are given hypothetical market scenarios and successfully generated quantitative freight index projection with confidence.

} 

\keywords{What-if Analysis, Multivariate Time Series Prediction, Human-centered Computing, Freight Rate Analysis}

\begin{document}


\firstsection{Introduction}

\maketitle

What-if analysis is crucial for decision-making under uncertainty. It answers the question of "What would happen to the model's prediction if input variable changes?" \cite{wexler_what-if_2019} While cross-sectional data sets are gaining functional support for user-defined variable perturbations, what-if capabilities for multivariate time series analysis tools are lagging behind.

We identity freight rate prediction as an exemplar task that requires what-if support: freight rates' extremely high volatility and tight co-integration factors mean that any slight perturbation to demand or supply factors bring about large changes to freight rate. This challenges freight analysts to keep an active eye out on market events and make informed and precise gauges about future freight rates using the "what-if" paradigm.

Literature on what-if tools for time series models is rare, two closest pieces related to the field include TimeFork\cite{badam_timefork_2016}, an interface to explore the inter-correlations between different time series variables. Google's What if tool\cite{wexler_what-if_2019} only allows editing and comparing cross-sectional data. To the best of our knowledge, there is no existing tool that enables users to easily manipulate time series input and understand the output of models in a comprehensive manner. 

We introduce a visual analytic solution to simulate market scenarios for freight rate prediction by probing the industry's commonly used multivariate time series models such as ARIMAX, multiple linear regression,VECM \cite{goulas_forecasting_2010} and LSTM\cite{naess_investigation_2018}. The system is created with four design goals in mind: intuitive manipulation of time series market data through line chart dragging; useful sensitivity analysis by generating scenario comparison views. Variable comprehension and model comparison that enable users to better understand the inner mechanism of models and spatial data aggregation for supply-side analysis. We've also conducted a case study with freight traders, the study shows that our solution can reliably assist users in anticipating and preparing for hypothetical market scenarios.

\section{Visual Design}
Our solution contains 6 views. Fig. 1a is the what-if input view which allows scenario adjustments. Fig. 1b is the coefficient impact view which facilitates indicator comprehension. Fig. 1c is the model comparison view that shows model performances for model selection and ranking. Fig. 1d is the spatial data view that supports aggregating spatial data. Fig. 1e is the prediction and what-if view. Fig. 1f is the what-if histories view which logs scenarios outputs.
\begin{figure}[htp]
    \centering
    \includegraphics[width=8cm]{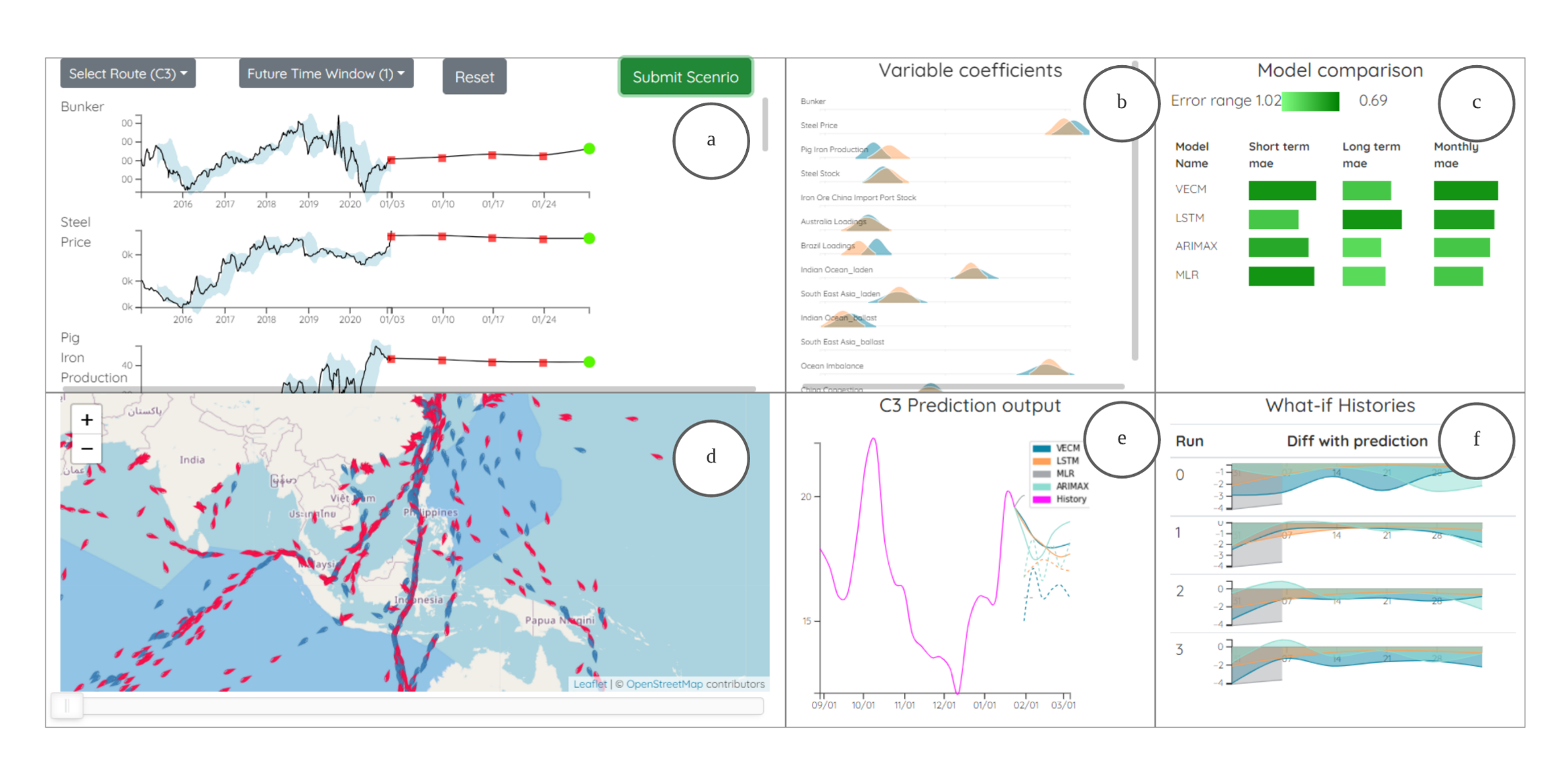}
    \caption{The what-if solution interface}
    \label{fig:galaxy}
\end{figure}

The what-if input view satisfies the first design goal, allowing users to flexibly create and update market scenarios. The historical line chart shows the general trend and seasonality of the variable data, it supports zooming through scrolling and displaying Bollinger Bands. The near term and the what-if data chart share the same time scale. The former magnifies the past 4 weeks' data. The latter applies a forward-fill strategy with the default forward window set to 1 to minimize the accumulation of error. Users can drag the green what-if data point vertically and the system updates the what-if value by inverse scaling the pixel shifts and variable values ranges. After dragging the data point(s) to the desired value(s), the user submits the data. The server computes the scenario output from time series models and the system displays the updated output in Fig. 1e and detailed comparison with the original prediction in Fig. 1f.

The variable impact factor view consists of vertically aligned area charts shaped in normal distribution. By showing their value range on the same chart, users get a grasp of the magnitude in which each variable affects the target time series variable. Thanks to the reduced height, the view also serves as an efficient variable localization window. Multiple regression-based and ARIMAX models have scalar variable coefficients for exogenous variables, we integrate them directly into the view. On the other hand, VECM models have vector coefficients and LSTM models hold model parameters in neurons and gates, we reached a design choice to not include them.


The spatial data aggregation view is a map displaying the raw spatial data and aggregated demand/supply-based information. Individual vessel information is represented by colored droplets. The tip of the droplets indicates the vessel's heading and the color reflects cargo status. More specifically, A red droplet highlights an empty(ballast) vessel whereas blue indicates the vessel is loaded with cargoes. Such information help traders identify potential vessels for chartering. Meanwhile, a unique vessel identifier number(IMO) is shown on hover when the user is interested in chartering the ship. In this way, users can make informed guesses about economic drivers with temporal features.

The what-if history view is a table, each row displays every run's what-if difference with the original predictions. It offers the user a complete history of scenario outputs. The X-Axis is positioned at the middle of the Y-Axis(Difference), its sign highlights the general trend of the freight index before and after applying the market scenario. The user can also inspect individual models' difference curves, by checking the curve's relative position around y=0, the user could have a grasp on the detailed trajectory of freight movement. This could prove useful in identifying hedging, speculation, and arbitrage opportunities.
\section{Evaluation}
We carried out a case study with three freight traders to assess the solution's usefulness. The process explored the effect of a torrential flood around a Brazil iron ore mine. The goal was to anticipate the resultant freight rate movement and uncover business opportunities for chartering. Each participant was tasked to perform the same set of actions on the interface as freight trader Alice.

\textbf{Prediction on original data and model preference: }Freight trader Alice observes that Minas Gerais, a state in Brazil with steady iron ore exports(freight demands), is experiencing an unusually heavy rainy season. She recalls the devastating Brumadinho dam disaster and its negative impacts on route C3's(from Brazil to China) freight rate\cite{vergilio_metal_2020}. Anticipating the possibility of another sharp drop in iron ore production in the region, she initiates a what-if analysis process to simulate the worst-case scenario. She first selects route C3 from Fig. 1a and waits for the pre-trained models' predictions based on the current week's market information Fig. 1e. She notices model VECM and LSTM enjoy significantly lower errors margins by looking at the model comparison view Fig. 1c. Anticipating the infrastructural damage and long-term consequences of the flood, she selects LSTM and VECM to be the primary models for the what-if scenario.

\textbf{What-if probing: }She starts looking for variables to change: Brazil loading and iron ore price. As the model has a fair amount of variables, she opted to look for it in the coefficient view as it encapsulates variables in a compact size. She clicked on "Brazil Loadings" in Fig. 2a, the input view automatically slides to the target row as shown in Fig. 2b. Then, looking at the past month's steady trends, she dragged down the production for 2000 mt as shown in Fig. 2c and clicks the "Submit Scenario" button.

\begin{figure}[htp]
    \centering
    \includegraphics[width=7cm]{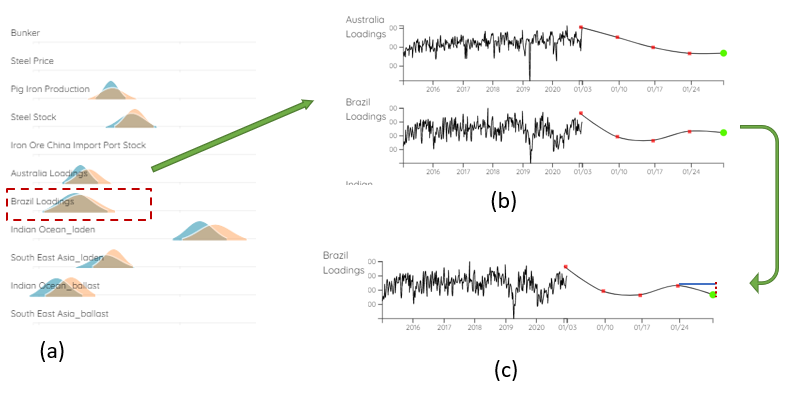}
    \caption{(a) User finds "loading" in the coefficient view and clicks. (b) The click event triggers what if window to scroll to loading's input section. (c) User changes the scenario value through dragging.}
    \label{fig:galaxy}
\end{figure}

\textbf{Anticipating scenario output and mitigate risks: }The what-if dotted lines in Fig. 3b go lower than the original prediction as anticipated. Meanwhile, she notices in the what-if history view that the middle line for the x-axis is nearly y=-2 in Fig. 3b. Now she has a clear outlook on what would happen should the rain continues. She starts looking for the ballast(empty) vessels approaching Brazil in the map view and taking down the IMOs of the vessels as shown in Fig. 4. She knows should the worst happen, some unfortunate ship owners will leave empty-handed from Brazil and be forced to charter out the vessels at a much lower price, that's when she'll strike for a bargain. 

\begin{figure}[htp]
    \centering
    \includegraphics[width=7cm]{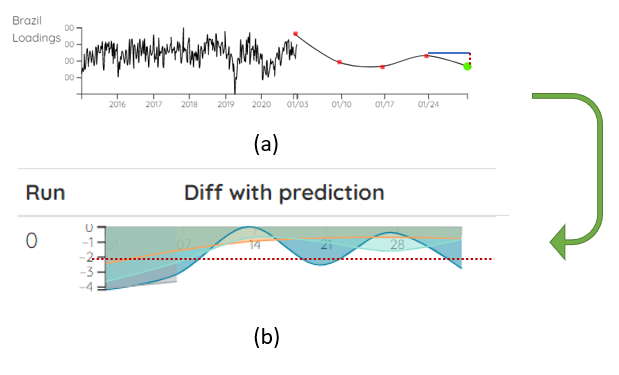}
    \caption{(a) Loading variable's new value. (b) The average difference between original prediction and updated scenarios is around -2.}
    \label{fig:galaxy}
\end{figure}

\begin{figure}[htp]
    \centering
    \includegraphics[width=7cm]{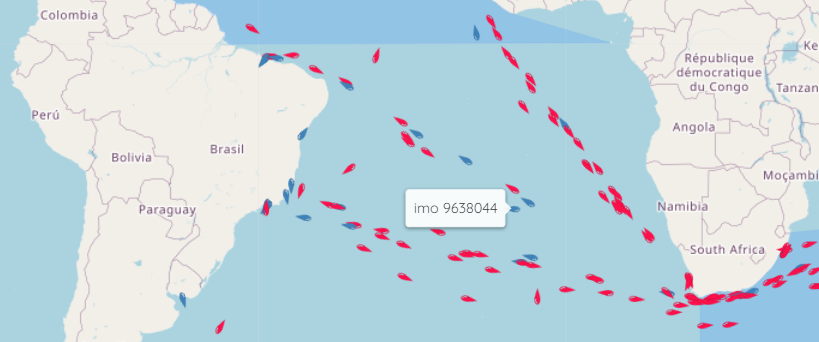}
    \caption{User locates empty vessels approaching Brazilian ports and identify potential chartering targets}
    \label{fig:galaxy}
\end{figure}

\section{Conclusion and future work}
We presented a what-if tool for manipulating and anticipating market scenarios for freight rate prediction. We evaluated the solution with industry practitioners. The user study reflects that the solution enables users to accurately capture freight index changes and respond to market events proactively. In the future, we plan to improve the accuracy of the machine-filled what-if data through learning statistical patterns of variables. We also aim to improve the applicability of the tool by incorporating seasonal parameters.


\bibliographystyle{abbrv-doi}

\bibliography{template}
\end{document}